\documentclass{aa}  

\usepackage{graphicx}
\usepackage{txfonts}
\usepackage{hyperref}
\usepackage{amssymb}
\usepackage{amsmath}
\usepackage{graphicx}
\usepackage{natbib}
\usepackage{rotating}
\usepackage{epsfig}
\usepackage{subfig}
\usepackage{colortbl}
\usepackage{lscape}
\usepackage{tabularx}
\usepackage{longtable}
\usepackage{color}

\usepackage{txfonts}
\newcommand{\civ}{C\,{\sc iv}~}

\newcommand{\mgii}{Mg\,{\sc ii}~}

\begin{document}

   \title{Restarting radio activity and dust emission in radio-loud\\ broad absorption line quasars}

  % \subtitle{I. Overviewing the $\kappa$-mechanism}

   \author{G. Bruni \inst{1,2}
 	 \and K.-H. Mack\inst{2}
	 \and F.M. Montenegro-Montes\inst{3}
	 \and M. Brienza \inst{4,5}
       \and J.I. Gonz\'alez-Serrano \inst{6}
}

   \institute{Max Planck Institute for Radio Astronomy, Auf dem H\"ugel 69, D-53121 Bonn, Germany
	 \and INAF - Istituto di Radioastronomia, Via P. Gobetti 101, I-40129 Bologna, Italy
	 \and European Southern Observatory, Alonso de C\'ordova 3107, Vitacura, Casilla 19001, Santiago de Chile, Chile
	 \and Netherlands Institute for Radio Astronomy, Postbus 2, 7990 AA, Dwingeloo, The Netherlands
       \and Kapteyn Astronomical Institute, Rijksuniversiteit Groningen, Landleven 12, 9747 AD Groningen, The Netherlands
       \and Instituto de F\'isica de Cantabria (CSIC-Universidad de Cantabria), Avda. de los Castros s/n, E-39005 Santander, Spain
		}					
							
   \date{}

% \abstract{}{}{}{}{} 
% 5 {} token are mandatory

  \abstract
  % context heading (optional)
  % {} leave it empty if necessary  
   {Broad absorption line quasars (BAL QSOs) are objects showing absorption from relativistic outflows, with velocities up to 0.2c. These manifest, in about 15\% of quasars, as absorption troughs on the blue side of UV emission lines, such as \civ and \mgii. The launching mechanism and duration of these outflows is not clear yet.}
  % aims heading (mandatory)
   {In this work, we complement the information collected in the cm band for our previously presented sample of radio loud BAL QSOs (25 objects with redshifts 1.7<z<3.6) with new observations at m and mm bands. Our aim is to verify the presence of old, extended radio components in the MHz range, and probe the emission of dust (linked to star formation) in the mm domain.}
  % methods heading (mandatory)
   {We observed 5 sources from our sample, already presenting hints of low-frequency emission, with the GMRT at 235 and 610 MHz. Other 17 sources (more than half the sample) were observed with bolometer cameras at IRAM-30m (MAMBO2, 250 GHz) and APEX (LABOCA and SABOCA, 350 and 850 GHz, respectively).}
  % results heading (mandatory)
   {All sources observed with the GMRT present extended emission at a scale of tens of kpc. In some cases these measurements allow us to identify a second component in the SED, at frequencies below 1.4 GHz, beyond the one already studied in the GHz domain. In the mm-band, only one source shows emission clearly ascribable to dust, detached from the synchrotron tail. Upper limits were obtained for the remaining targets.}
  % conclusions heading (optional), leave it empty if necessary 
   {These findings confirm that BAL QSOs can also be present in old radio sources, or even in restarting ones, where favourable conditions for the outflow launching/acceleration are present. A suggestion that these outflows could be precursors of the jet comes from the fact that $\sim$70\% of our sample can be considered in a GigaHertz Peaked Spectrum (GPS) or Compact Steep Spectrum (CSS)+GPS phase. This would confirm the idea proposed by other authors that these outflows could be recollimated to form the jet. Comparing with previous works in the literature, dust emission seems to be weaker than the what expected in `normal' QSOs (both radio loud and radio quiet ones) suggesting that a feedback mechanism could inhibit star formation in radio-loud BAL QSOs.}

   \maketitle
%
%________________________________________________________________

\section{Introduction}

In the context of Active Galactic Nuclei (AGN) outflows from the central region are commonly detected as absorption lines in more bands. In the UV/Optical range, they can be present in $\sim$70\% of type 1 AGN, with an extension up to kpc scales and velocities up to $\sim$1000 km/s (\citealt{Harrison}). A similar fraction ($\sim$60\%) has been found in the quasar (QSO) population (\citealt{Gan}).
In the X-ray band, more highly ionized outflows are detected as Ultra Fast Outflows (UFO) both in radio quiet (\citealt{Pounds}) and radio loud AGN (\citealt{Tombesi14}) with much higher velocities, in the range 0.03-0.4c.
A feedback effect from outflows on the host galaxy has been proven to exist by different authors in the past years (e.g. \citealt{Feruglio,Wang,Sturm}), and lately they have been proven to hamper star formation (\citealt{Tombesi}).

BAL QSOs are among the objects presenting the fastest outflows. These are detected in about 15\% of QSOs as broad absorption troughs in the UV spectrum, on the blue side of emission lines from ionised species, mainly \civ and \mgii. They can be both detached or superimposed to the emission peak, and reach relativistic velocities of up to 0.2c (\citealt{Hewett}). \cite{Allen} have found a dependence with redshift of the BAL fraction, decreasing a factor of 3.5$\pm$0.4 from z$\sim$4 to $\sim$2.
The mechanism at the origin of these violent outflows has not been unveiled yet. The two main scenarios discussed in the literature tend to ascribe the BAL phenomenon to: 1) young objects, in which the strong nuclear starburst activity is still expelling a dust cocoon (\citealt{Briggs, Sanders, Farrah}), or 2) normal QSOs, whose outflows intercept the line of sight of the observer (\citealt{Elvis}). In this case, relativistic outflows are supposed to be commonly present in QSOs, but detected only when orientation is favourable. Variability of the BAL troughs has been explored by many authors in the past years, thanks to the increasing amount of available spectroscopic surveys data (\citealt{Gibson08,Gibson10,Capellupo11,Capellupo12,Vivek12}), and a typical duty cycle of about a thousand years for the BAL-producing outflow has been found (\citealt{Filiz12,Filiz13}).

Several works have been published in the past years, trying to collect information in the different electromagnetic bands. In particular, the emission in the radio band has been used to probe the orientation and age of these objects (\citealt{Montenegro,DiPompeo1,Bruni,Bruni2}).
No clear hints of a favoured scenario arose from these works, resulting in indications of different possible orientations and different ages for radio loud (RL) BAL QSOs. In this work, we present follow-up observations of sources from our previously studied sample (\citealt{Bruni}). We explored the emission properties at m- and mm-wavelengths, to complement the multi-wavelength view of these objects, already studied at cm-band in our previous work.

The detection of a strong MHz emission can be safely interpreted as the presence of old extended radio plasma connected to a former AGN radio-active phase, that can be as old as 10$^7$-10$^8$ years (\citealt{Konar06,Konar13}). It has been shown indeed that jets in radio loud AGN can have multiple phases of activity (\citealt{Lara,Schoen,Saikia,Nandi}) with a duty cycle that depends on the source radio power (\citealt{Best2,Shabala}). Hints of these components in BAL QSOs were already found from literature data, for the same sample studied in this paper, and presented in \cite{Bruni}. This kind of emission adds a significant information in the framework of the presented models for BAL QSOs. In particular, if the young-scenario would be the most realistic one, no further components than the one peaking in the MHz-GHz range should be present, since GigaHertz-Peaked Spectrum sources (GPS) and Compact Steep Spectrum sources (CSS, \citealt{ODea}), together with High Frequency Peakers (HFP, \citealt{Dallacasa}) are among the youngest radio sources. To date systematic searches of diffuse emission around GPS and CSS sources find about 20\% detections (\citealt{Stan1,Stan2}). In light of this we use low-frequency GMRT observations to probe possible extended emission around BAL QSO with the aim of possibly confirming or discarding the youth scenario described in the previous paragraph. The radio phase itself does not seem to introduce significant differences in RL BAL QSOs with respect to Radio Quiet (RQ) ones (\citealt{Bruni3,Rochais}), thus being a valid tool to study general phenomenology of these objects.

The continuum emission of dust, in the rest-frame far-infrared domain (FIR), can be detected at mm-wavelengths (over 100 GHz) for objects with z$\sim$2. Objects enshrouded by gas and dust can host star-formation regions (\citealt{Zahid}), and thus show high star-formation rates, that may indicate a young age of the galaxy. A flux density excess in the FIR could be an indicator of a different age for BAL QSOs with respect to the non-BAL QSO population, and thus help in discriminating between the orientation and the evolutionary models. There are two major works presenting (sub)mm observations on samples of BAL QSOs. \cite{Willott} showed SCUBA measurements on a sample of 30 radio-quiet
BAL QSOs and conclude that there is no difference between BAL QSOs and a
comparison sample of non-BAL QSOs. Nevertheless, \cite{Priddey} based on SCUBA observations of 15 BAL QSOs found tentative evidence for a dependence of submm flux densities on the equivalent width of the characteristic \civ BAL which \emph{`suggests that the BAL phenomenon is not a simple geometric effect [\dots] but that other variables, such as evolutionary phase, [\ldots] must be invoked'}. \cite{Cao} discuss the far-infrared properties and star-formation rates (SFR) of BAL QSOs (without distinguishing between RL and RQ ones), using data from the {\emph{Herschel}}-ATLAS project. They found no differences with respect to non-BAL QSOs, concluding that a scenario in which BAL QSOs are objects expelling a dust cocoon is improbable.
 
The main difference of the above samples compared to our target sample is 
the radio-loudness of our sources. 
With the radio-data presented in \cite{Bruni} we were able to characterise the synchrotron
spectra of our sources, and thus to study the peak frequency and spectral index distributions with respect to the `normal' QSO population. Also an upper limit to the synchrotron emission at mm-wavelengths can be constrained.

A search for HFP, i.e. the
youngest known radio sources with the highest turnover frequencies, shows 
only a very small percentage of sources with peak frequencies close to 20 GHz
(\citealt{Dallacasa}) with the most extreme case at 25 GHz, leading to a 
formal age of some 50 years only (\citealt{Orienti}).
As any upturn towards an even higher peak frequency would be visible in our Spectral Energy Distributions
(SEDs) we can safely assume that the extrapolated synchrotron emission
reflects its true contribution at 250 GHz and that any observed excess emission
can be attributed to the presence of cold dust. Moreover, the variability study presented in \cite{Bruni} excludes any possible significant variability even at high frequencies over a 3 years time scale, for this sample of objects.

The outline of the paper is as follows: in Sect.~\ref{sec:sample} we describe the BAL QSO sample. The radio observations are reported in Sect.~\ref{sec:observations}. In Sect.~\ref{sec:results} we present the results concerning morphology at MHz frequencies and dust abundance. Sect.~\ref{sec:discussion} is a discussion of the results, in the context of recent works about BAL QSOs.

The cosmology adopted throughout the work assumes a flat universe and the following parameters: $H_0=71$ km s$^{-1}$ Mpc$^{-1}$, $\Omega_{\Lambda}=0.73$, $\Omega_{M}=0.27$.

\section{The RL BAL QSO sample}
\label{sec:sample}
	The radio-loud BAL QSO sample studied in this paper is presented in \cite{Bruni}. All sources were chosen among objects from the 4th edition of SDSS Quasar Catalogue (\citealt{Schneider07}), drawn from the 5th data release of the Sloan Digital Sky Survey (SDSS-DR5; \citealt{Adelman}). To select radio-loud objects, we cross-matched the SDSS with the FIRST (Faint Images of the Radio Sky at Twenty-cm; \citealt{Becker2}) and only those with a counterpart lying $<$2 arcsec away and having $S_{1.4\rm{GHz}}$ $>$ 30 mJy were considered. All of these satisfy the radio-loudness definition by \cite{Stocke}. Moreover, the selection has been limited to those objects whose redshifts lie in the range 1.7 $<z<$ 4.7 allowing the identification of both \civ and \mgii absorption features on SDSS spectra. In order to select genuine BAL QSOs, only objects with an Absorption Index (AI) $>$ 100 $\rm{km/s}$ were considered, and only troughs broader than 1000 $\rm{km/s}$ were used for this calculation\footnote{i.e. we adopted an AI defined as ${\rm{AI}}=\int_{0}^{25000}(1-\frac{f(v)}{0.9})\cdot Cdv$, as in \citealt{Hall}, but with $C$=1 only for contiguous troughs $\ge$ 1000 km/s, and null otherwise.}. This resulted in 25 radio-loud BAL QSOs. For a complete description of the sample and the selection procedure refer to \cite{Bruni}.

\begin{table*}[]
	\caption{Summary of the observations and setups presented in this paper. IRAM and APEX telescopes made use of bolometer receivers.}
\begin{center}
%\small
%	\centering
	\scalebox{0.92}{
	\begin{tabular}{clccccc}
		\hline
		\hline
			Run & Date &	Telescope	& Frequency & Bandwidth  & FWHM & Number of sources\\
			&&& (GHz) & (MHz) & (arcsec) & \\
			\hline
1&			9-11 Jan 2010 					& GMRT 			& 0.235, 0.610		&	33  	&	15-45, 4-10	&	5	\\
2&			26 Oct-23 Nov 2010 				& IRAM-30m 		& 250			&	-	&	11			&	11	\\			
3&			05-22 Aug 2010				& APEX			& 850			&	-	&	19, 8			&	4	\\
4&			01-09 Sep, 05 Nov 2011 			& APEX			& 345, 850		&	-	&	19, 8			&	2	\\
5&			04-05 Jun, 13-14 Aug 2012 		& APEX 			& 850			&	-	&	19, 8			&	3	\\
			\hline
			\end{tabular}}
	\label{Summaryofobservations}
\end{center}
\end{table*}

\section{Radio observations and data reduction}
\label{sec:observations}

In this paper, we present observations complementary to the ones performed by \cite{Bruni} at cm wavelengths. The GMRT, the APEX single-dish and the IRAM 30-m telescope were used to extend the available SEDs extension.
Table ~\ref{Summaryofobservations} summarises the different runs and observing setups.

\subsection{Giant Metrewave Radio Telescope}

Observations at frequencies of 235 MHz and 610 MHz with the GMRT were performed for 
five sources, during January 2010. We used the double frequency mode to observe 
simultaneously at the two frequencies. The total bandwidth 
for each band was 33 MHz, divided into 256 channels of 0.13 MHz each.
We observed in snapshot mode to improve the UV coverage for the sources.
Standard phase and amplitude calibration were performed, using 3C 286 as
primary calibrator every $\sim$4 hours and suitable phase calibrators near targets 
every $\sim$30 minutes. Correlation was performed using the GSB software correlator at NCRA.
Data were reduced with the AIPS\footnote{http://www.aips.nrao.edu/index.shtml} package, using the standard procedures.
Flux densities were extracted from images via Gaussian fit of the components, using task JMFIT inside AIPS.

\subsection{IRAM-30m single dish}

We could observe 11 sources of the BAL QSO sample at 250 GHz with the IRAM-30m telescope, during the 2010 summer pool session. We used the MAMBO2 117-pixel bolometer in ON/OFF mode, since all of our sources are point-like for this telescope (HPBW=11 arcsec).
With average atmospheric conditions the detector could reach a noise of $\sim$1 mJy/beam in $\sim$40 min of observing time: we observed each source for this duration in order to obtain a detection or a 3-$\sigma$ upper limit.
Skydip, calibration and pointing scans were regularly performed during the runs and every time the observing direction in the sky significantly changed in elevation. Focus was repeated at sunrise and sunset. A standard reduction was done using the MOPSIC\footnote{http://www.iram.es/IRAMES/mainWiki/CookbookMopsic} script provided by IRAM.

\subsection{APEX}

From 2010 to 2012 we observed with APEX a total of 9 southern sources from the BAL QSO sample.
All of them were observed with the SABOCA bolometer array (\citealt{Siringo2}) 
at 850 GHz in photometry mode (HPBW $\sim$8 arcsec) and two of the sources (0044+00 and 1404+07) also 
in mapping mode. %Project ids for these observations are 085.B-0354A, 088.F-9308A, and 089.F-9307A. 
In addition, one of the sources (0044+00) was observed with the LABOCA bolometer array 
(\citealt{Siringo1}) at 345 GHz in photometry mode (HPBW $\sim$19 arcsec). % under project 088.F-9308B. 
APEX observations were carried out in service mode, with typical integration times 
of ~1 h per source, in order to reach RMS values around 20 mJy/beam. Calibration was based on 
observations of primary calibrators (Mars and Uranus) as well as skydips measured at the same 
azimuth of the targets to derive atmospheric opacity. A standard reduction was done using the 
version 2.15-1 of the CRUSH\footnote{http://www.submm.caltech.edu/~sharc/crush/} software which offers an improved pipeline for photometric data as compared to earlier versions.

\subsection{Error determination}

In the flux density error calculation, different contributions were 
considered for the GMRT interferometric observations:

\begin{itemize}
\item The thermal noise, $\Delta S _{\rm{noise}}$, which is estimated from the map, in empty regions of sky surrounding the target;\\
\item The fractional calibration error, $\Delta S_{\rm{calib}}$, estimated as the visibilities dispersion of the flux density calibrators;\\
\end{itemize}

In particular, we followed the approach proposed by Klein et al. (2003). The expression used is reported below:
\begin{equation}
\Delta I=\sqrt{(\Delta S_{\rm{calib}}\cdot S)^2 +(\Delta S _{\rm{noise}})^2 \cdot \frac{A_{\rm{src}}}{A_{\rm{beam}}}},
\end{equation}
where $A_{\rm{beam}}$ and $A_{\rm{src}}$ are respectively the area of the synthesised beam and the aperture used to extract the source flux density. From their ratio we determine the number of beams contained in the source. 

For APEX and IRAM-30m data, obtained with bolometer receivers, the error was calculated using the respective packages for data reduction, estimating the noise from off-source subscans.

\subsection{The polarimetry campaign}

During 2011, we conducted a polarimetry campaign using the EVLA and the Effelsberg-100m single dish on this same sample, to implement data later presented in \cite{Bruni}, and probe with deeper observations the polarisation of the faintest sources. The results from this campaign will be presented in a future paper, but we decided to use part of the obtained total flux density measurements to improve the SED coverage of this work (see Tab. \ref{errata}). Also, one of the sources observed with the EVLA turns out to have a resolved structure (0849+27), and we present here the map (see Sect. \ref{sec:morphology}). This same source was also observed during our mm campaign (see Tab. \ref{dust_flux}). Observations and data reduction were conducted as in \cite{Bruni}.
\begin{table}
\centering
\caption{Top lines: revised flux densities for sources presented in \cite{Bruni} and in this work. Bottom lines: flux densities from our polarimetry campaign, used for this work.}
\begin{tabular}{cccc}
\hline
\hline
ID & Frequency & S & Telescope \\
(J2000)   &  [GHz]         & [mJy] & \\
\hline
0756+37 & 43  &  5.2$\pm$0.8 		& VLA \\
0816+48 & 1.4 &  70.9$\pm$0.7 	& VLA \\
1335+02 & 43  &  8.4$\pm$1.3 		& VLA \\
\hline
0842+06 & 8.35 & 19.5$\pm$0.8 	& Effelsberg-100m \\
0849+27 & 4.86 & 27.6$\pm$0.7 	& EVLA \\
\hline
\end{tabular}
\label{errata}
\end{table}
%%%%%%%%%%%%%%%%%%%%%%%%%%%%%%%%%%%%%%%%%%%%%%%%%%%%%%%%%%%%%%%%%%%%%
%
%
\begin{figure*}
\begin{center}
	\includegraphics[width=18cm]{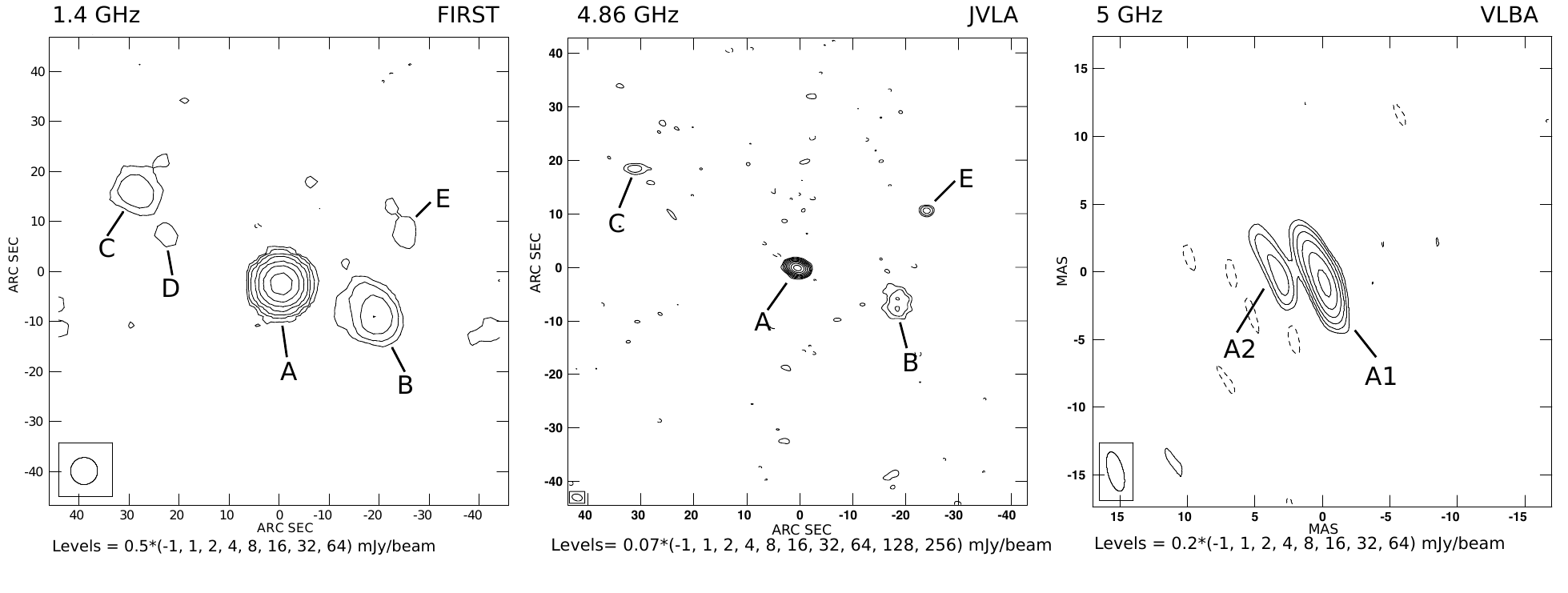}
	\caption{Maps of BAL QSO 0849+27 from the FIRST survey (left panel - 1.4 GHz, beam 5.40$\times$5.40 arcsec), our EVLA observations (central panel - 4.86 GHz, beam 1.95$\times$1.25 arcsec), and our VLBA observations (right panel - 5 GHz, beam 2.99$\times$1.15 mas). Contours are multiples of 3-$\sigma$, according to label. Dashed contours are negative. The synthesised beam size is shown in the lower left corner of each map. Note that the scale of the right panel is in mas.}
\label{0849}
\end{center}
\end{figure*}

\section{Results}
\label{sec:results}

In the following, we present the results from our observing campaign: the morphology could be studied with the GMRT interferometer, while images from the IRAM and APEX bolometers were used for photometric measurements. The collected information is presented in Tab. \ref{GMRT_size}, while SEDs, including flux densities at cm wavelengths from \cite{Bruni} and our polarimetry campaign, are presented in Fig. \ref{dust} and discussed in Sect. \ref{sec:SED_section}. 

%%%%%%%%%%%%%%%%%%%%%%%%%%%%%%%%%%%%%%%%%%%%%%%%%%%%%%%%%%%%%%%%%%%%%%%%%%%%%%%%%%

\subsection{Morphology}
\label{sec:morphology}

From the GMRT and EVLA maps we were able to investigate the morphologies of the sources at arcsec scales.
The frequency range explored with the GMRT allowed us to put some constraints to
the presence of extended, old, radio components.

\subsubsection*{GMRT maps}

\begin{figure*}
\begin{center}
	\includegraphics[width=18cm]{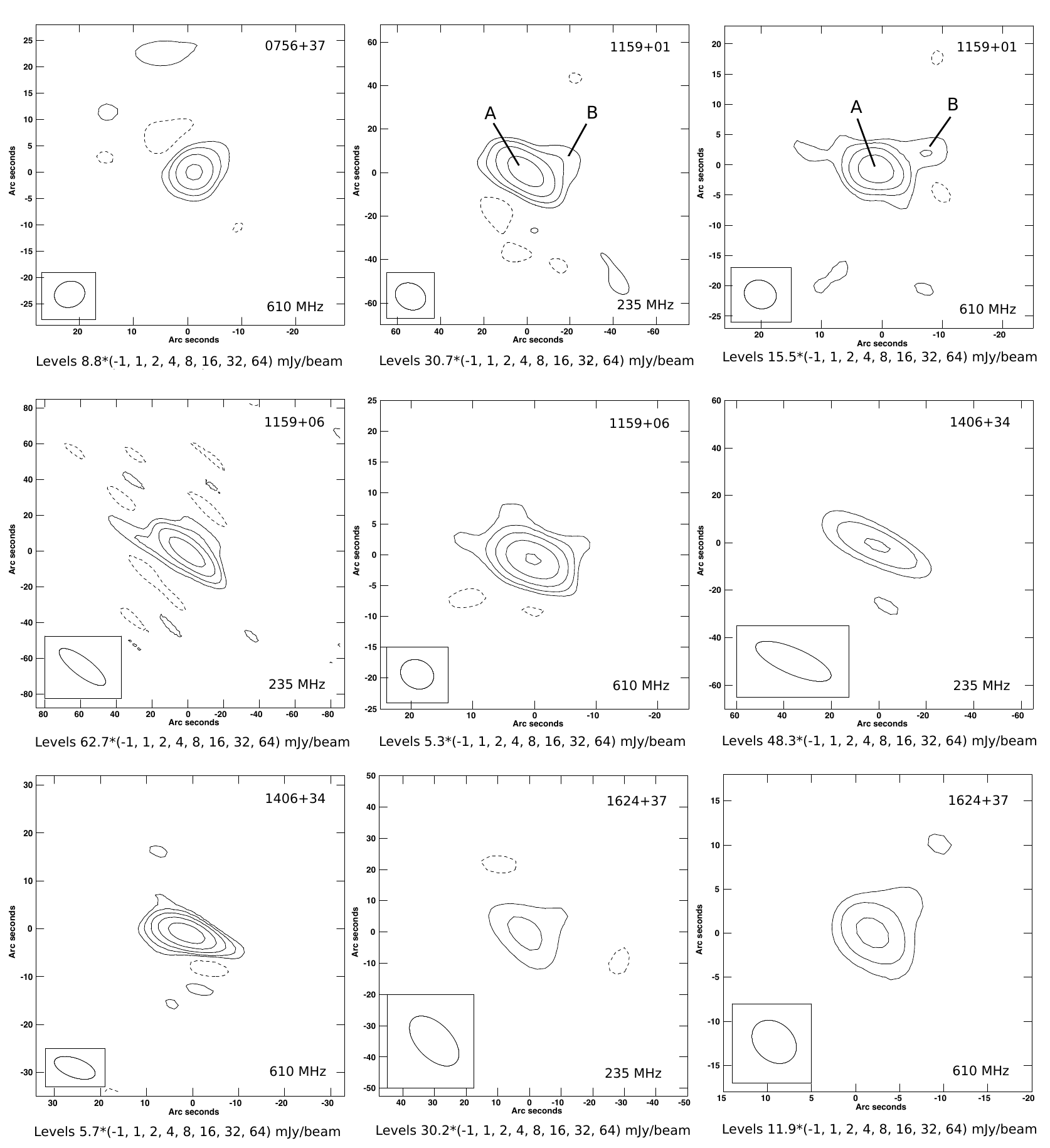}
	\caption{Maps of 5 BAL QSOs observed with the GMRT at 235 and 610 MHz. Contours are multiples of 3-$\sigma$, according to the label. Dashed contours are negative. The synthesised beam size is shown in the lower left corner of each map.}
\label{GMRT}
\end{center}
\end{figure*}

We observed with the GMRT the five sources from our BAL QSO sample showing the strongest low-frequency 
emission in the flux densities collected from archival surveys data. The goal was the detection of extended emission at 235 or 610 MHz, indicating a previous radio-activity period of the central AGN, and thus putting a constraint on the age. Maps show components with deconvolved dimensions greater than zero in most cases, corresponding to a fraction of the beam (see Fig. \ref{GMRT}). 
One source (1159+01) presents an elongated structure at 235 MHz, confirmed by the detection of a second component (B) at higher resolution in the 610 MHz map. This structure is compatible with the one seen in the pc-scale maps obtained by \cite{Tak}, where a jet extension at a comparable position angle is visible. This could confirm the presence of two different radio phases, at different scales, as also highlighted by the SED of this object (see Sec. \ref{sec:SED_section}). Quantities for all sources at the two frequencies are presented in Table \ref{GMRT_size}, together with projected linear sizes. 

These measurements confirm the presence of a low-frequency, older radio component in some BAL QSOs,
thus excluding that they are a subclass of young radio objects. The size of these components is significantly larger
than the values of a few kpc measured for the high-frequency, unresolved, counterparts (see \citealt{Bruni}), thus suggesting different emitting regions for the two. The flux densities found nicely fit with collected data from surveys (see Fig. \ref{dust}).

\begin{table*}[]
\renewcommand\tabcolsep{3.0pt}
	\caption{Information collected with the GMRT interferometer: Col. 2 and 8 are flux densities from the maps, Col. 3-6 and 9-12 are deconvolved major and minor axes at 235 and 610 MHz, respectively. The projected linear size in kpc is also given. When a null deconvolved size were found, we considered the corresponding beam size as upper limit. For the only resolved source (1159+01) we give values for both components.}
	\label{GMRT_size}
	\centering
		\begin{tabular}{cccccccccccc}
		\hline
		\hline
		
		& \multicolumn{5}{c}{235 MHz} 
		& \multicolumn{5}{c}{610 MHz} \\

		ID 	& $S$ &Maj. axis   & Min. axis & Maj axis & Min. axis& Component& $S$ &Maj. axis   & Min. axis & Maj axis & Min. axis\\
			& (mJy)	& (arcsec)	& (arcsec)	& (kpc)		& (kpc)   && (mJy)	& (arcsec)	& (arcsec)	& (kpc)		& (kpc) \\
		\hline
0756+37 	&	$<$85.2			& 	- 			& - 			& - 			& - 			&	&112$\pm$4		& 3.9&2.2&32.4&18.3\\
1159+01 	&	720$\pm$65		&	19.6			&7.0			&164~~~~		&58.4		&A	&~~312$\pm$10	&4.6&2.3&39.5&19.8\\
		&					&				&			&			&			&B	&~~74.8$\pm$4.2		&10.2~~~&0.9&86.3&~~7.6\\
1159+06 	&	872$\pm$81		&	~~6.9		&$<$10.1		&58.7		&$<$85.9		&	&163$\pm$5		&6.7&2.4&57.0&20.4\\
1406+34 	&	249$\pm$26		&	14.8			&$<$11.0		&122~~~~		&$<$90.9		&	&168$\pm$5		&3.9&1.1&32.2&~~9.1\\
1624+37 	&	104$\pm$16		&	11.1			&$<$11.0		&84.9		&$<$84.1		&	&~~98.6$\pm$4.9	&4.3&3.7&32.9&28.3\\
		\hline	
		\end{tabular}
\end{table*}

\subsubsection*{EVLA map of 0849+27}

During our polarimetry campaign, we could observe the peculiar BAL QSO 0849+27 with the EVLA at 4.86 GHz. The map of the resolved structure of this source at 1.4 GHz, obtained from the FIRST survey, was already presented in a previous work from our group (\citealt{Bruni}). This resulted to be the most extended BAL QSO of our sample (44 arcsec, 382 kpc, between components A and C). We could obtain a map at 4.8 GHz from our subsequent EVLA observation in 2011, taking advantage of the improved performance of this instrument, and also a high resolution (pc-scale) map fo the core component from our VLBA program (see \citealt{Bruni2}). Results are presented in Fig. \ref{0849}. Four out of five components detected in the FIRST data are visible in our EVLA map (A, B, C, E), while the flux density of component D seems to drop below the 3-$\sigma$ significance level. Total flux densities as measured at 4.86 GHz are 27.6$\pm$0.7 mJy, 2.0$\pm$0.2 mJy, 0.49$\pm$0.05 mJy, and 0.41$\pm$0.05 mJy for component A, B, C, E, respectively. Components E was not classified in our previous FIRST map, but given the clear detection we obtained in the EVLA map, we extracted the flux density at the corresponding position in the FIRST map, where a single contour was present. This results to be 2.36$\pm$0.15 mJy. 
The obtained spectral index for these components, are $-$0.54$\pm$0.04,  $-$1.10$\pm$0.17, $-$1.78$\pm$0.18, and $-$1.41$\pm$0.22 for components A, B, C, and E, respectively. A flat spectral index ($>-$0.5) usually identifies the core for non-Doppler-boosted components: component A shows a spectral index compatible with that value within the error, while B, C, and E have a steep spectral index ($<-$0.5). From our VLBA observations we could confirm that core is component A, since at pc-scale it shows a core-jet structure, with one of the two components (A1) having a flat spectral index of $-$0.12$\pm$0.20 between 5 and 8.4 GHz.

The peculiar morphology of this source, explored at both kpc and pc scales, suggests a jet precession, or radio-activity phases with different jet axes. In fact, trajectories connecting the core (A) with the other components detected at kpc scale are all different, not permitting to identify one as the counter-jet of the other. Moreover, the pc-scale structure shows a further launching direction for the most recent component (A2), not corresponding to any among the ones at kpc scale. In this scenario, the BAL-producing outflow would be present in a source presenting multiple ongoing radio phases: this would not be easily explicable with the young scenario.

%%%%%%%%%%%%%%%%%%%%%%%%%%%%%%%%%%%%%%%%%%%%%%%%%%%%%%%%%%%%%%%%%%%%%%%%%%%%%%%%%%

%%%%%%%%%%%%%%%%%%%%%%%%%%%%%%%%%%%%%%%%%%%%%%%%%%%%%%%%%%%%%%%%%%%%%%%%%%%%%%%%%%

\subsection{SEDs shape from 235 MHz to 850 GHz}
\label{sec:SED_section}

We present here a study of the SED shape for the objects in this work. We implemented the new flux densities 
in the ones from \cite{Bruni}, spanning from 74 MHz up to 43 GHz, in order to improve the overall frequency coverage. 
For three sources (0756+37, 0816+48, 1335+02) we provide here a revised flux density for the VLA measurements presented in \cite{Bruni}: for source 0816+48, during previous data reduction the vicinity of a strong source led to an incorrect phase referencing, resulting in an overestimated flux density measurement. While for sources 0756+37 and 1335+02, our flux-extraction algorithm missed the source, displaced of a few arcsec from the map centre, due to atmospheric effects. The revised values are given in Tab. \ref{errata}.

\subsubsection*{Flux densities at mm wavelengths}

Flux density from the dust grey-body thermal emission can be described by the following equation (\citealt{Hughes}):
\begin{equation}
S^{\rm{obs}}= \frac{(1+z)}{D_L^2} \times M_dk_d^{\rm{rest}}B(\nu^{\rm{rest}},T_d),
\label{Sdust}
\end{equation}
where $S^{\rm{obs}}$ is the observed flux density at a given frequency $\nu_{obs}$, and $\nu_{rest}$ is the rest-frame frequency, $D_L$ is the luminosity distance, $z$ is the target
redshift, $k_d^{\rm{rest}}$ is the mass absorption coefficient at a given rest-frequency, $B$ is the black-body
Planck function, $T_d$ is the dust temperature, and $M_d$ is the dust mass. Typical AGN values for $T_d$ and $M_d$ from the latest works in the literature
are 10<$T_d$<60 K, and $M_d\sim10^8 M_\odot$ (\citealt{Kal}), while $k_d^{\rm{rest}}$ is usually calculated scaling down to the desired wavelength values estimated in previous works (e.g. $k_d^{\rm{850\rm{\mu m}}}$=0.077 m$^2$kg$^{-1}$ in \citealt{Dunne00,Dunne11}), approximating the trend \emph{vs} wavelength as a power law: 
\begin{equation}
k_d^{\rm{rest}}\propto\lambda^\beta, 
\end{equation}
where $\beta$ is the dust emissivity index (\citealt{daCunha}). 
We could collect flux densities at mm-wavelengths for a total of 17 objects from the BAL QSO sample. 
In most cases only upper limits could be derived (see Tab. \ref{dust_flux}), nevertheless providing useful constraints for the dust abundance. Only one source (0756+37) shows a flux density at 250 GHz clearly distinguishable from the expected synchrotron-emission tail: this is the only genuine example of dusty BAL QSO found among these objects. 
Despite the fact that we could not collect enough detections in the mm-band to perform a full fit of the dust emission using Eq. \ref{Sdust}, we can still compare the obtained upper limits with the results from previous works in the literature (\citealt{Omont,Kal}, see Sec. \ref{sec:discussion:dust} for a discussion).

\subsubsection*{SED fitting}

\begin{figure*}
\begin{center}
	\includegraphics[width=18.5cm]{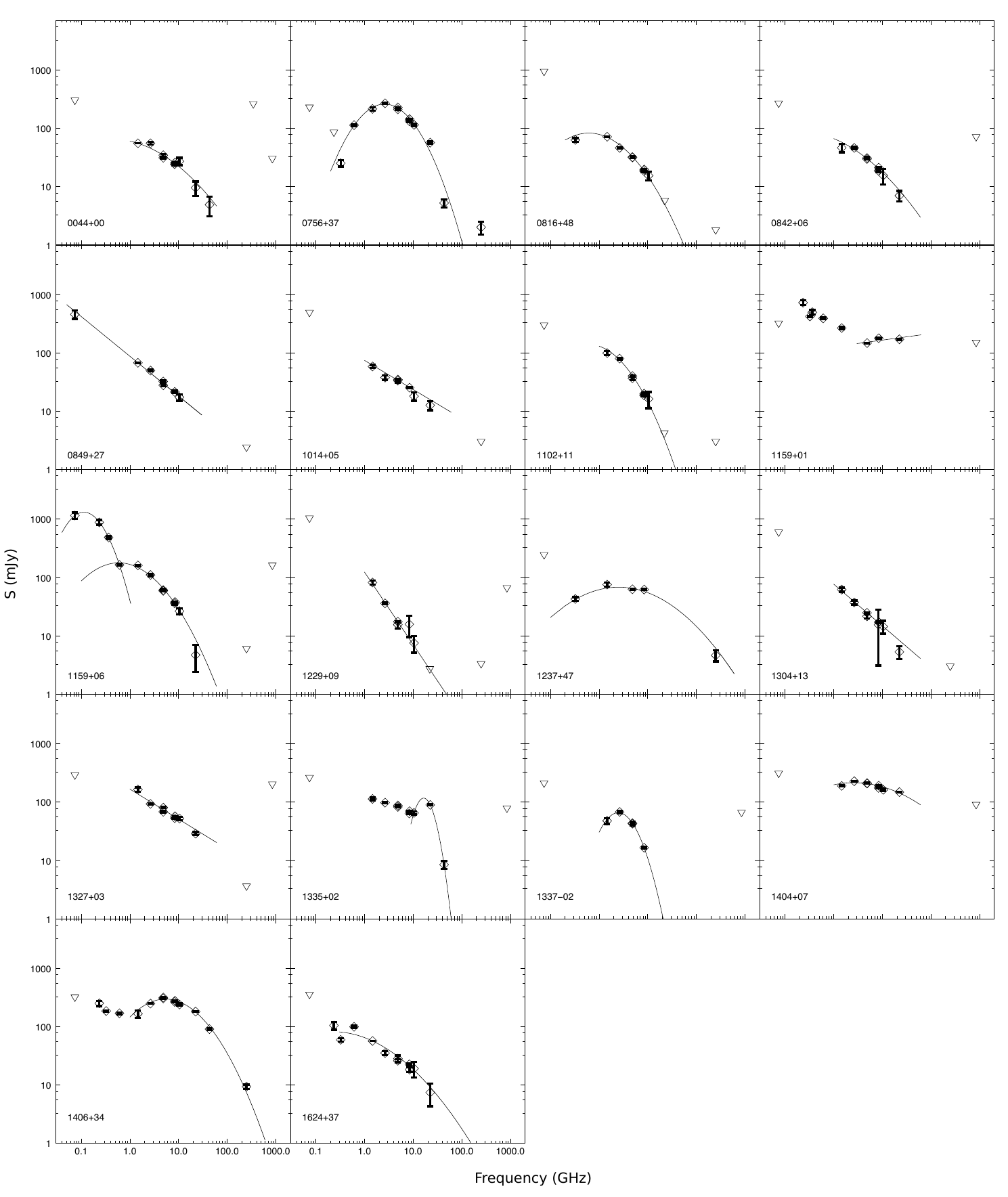}
	\caption{SEDs of the 18 BAL QSOs observed during the m/mm-wavelengths campaign (x-axis: GHz; y-axis: mJy). 235 MHz and 610 MHz flux densities from the GMRT; 250 GHz flux densities from IRAM-30m; 345 and 850 GHz flux densities from APEX. Measurements at other frequencies are taken from \cite{Bruni}, or from our polarimetry campaign (see Sect. 3). Triangles are 3-$\sigma$ upper limits. Solid lines are parabolic or linear fits, according to the criteria discussed in Sec. 4.2.}
\label{dust}
\end{center}
\end{figure*}

For the SED fitting in the m-/cm-wavelengths domain, we adopted the same basic approach as in our previous work, i.e. a linear and a parabolic fit in logarithmic scale, determining which of the two functions fits best the data. The former is a simple model of power-law synchrotron emission, commonly used to fit the optically-thin part of the emission above the peak, for old sources peaking at frequencies below the available ones. The latter is a first-order approximation of an optically-thick (on the left side of the peak) plus an optically-thin synchrotron emission (on the right side of the peak) capable of fitting the SED of a young radio source, peaking in the MHz-GHz range.
Upper limits were excluded from the dataset used for fitting, as well as datapoints clearly belonging to a second component in the MHz range (if any). After introducing the new data, we obtained best fits with the same functions (line or parabola) than previously adopted, except for sources 0842+06, that resulted to have a parabolic shape, and 1335+02, that after the introduction of the revised flux density from Tab. \ref{errata} show a HFP component peaking at $\sim$20 GHz (see Fig. \ref{dust}). Another source (1159+06) showed a second synchrotron component in the MHz range, peaking at about 100 MHz, while 0816+48, that was not fitted in \cite{Bruni}, now presents a parabolic component in the GHz range using again the revised flux density at 1.4 GHz.

In two of the three sources detected at 250 GHz the emission is most probably due to synchrotron component (1237+47, 1406+34), since it perfectly fits in the shape of the emission tail. For source 0756+37 we found a significant excess of emission at 250 GHz (2.0$\pm$0.5 mJy), with respect to the expected contribution of the synchrotron emission at the same frequency. Thus we can consider this emission as most probably produced by dust.

%%%%%%%%%%%%%%%%%%%%%%%%%%%%%%%%%%%%%%%%%%%%%%%%%%%%%%%%%%%%%%%%%%%%%%%%%%%%%%%%%%

\begin{table}
%\vspace{16cm}
%\begin{minipage}{250mm}
\centering
\caption{Results for the 17 BAL QSOs observed in the mm-band: 250 GHz flux densities from IRAM-30m, 345 and 850 GHz from APEX. We give 3-$\sigma$ upper limits in case of non-detection.}
\label{dust_flux}
\renewcommand\tabcolsep{5pt}
\begin{tabular}{lccccc}
\hline
\hline
\multicolumn{1}{c}{Name}       	&
%\multicolumn{1}{c}{$S_{0.235}$}  	& 
%\multicolumn{1}{c}{$S_{0.610}$}  	&
\multicolumn{1}{c}{$S_{250}$}  	& 
\multicolumn{1}{c}{$S_{345}$}  	&
\multicolumn{1}{c}{$S_{850}$} 	\\

\multicolumn{1}{c}{}       	&
%\multicolumn{1}{c}{(mJy)}  	& 
%\multicolumn{1}{c}{(mJy)}  	& 
\multicolumn{1}{c}{(mJy)}  	& 
\multicolumn{1}{c}{(mJy)}  	& 
\multicolumn{1}{c}{(mJy)} 	\\
\hline
0044+00	&	-				&$<$260 		&$<$30	\\
0756+37	&	2.0$\pm$0.5 	&-				&- 			\\
0816+48	&	$<$1.8 			&- 				&- 			\\
0842+06	&	-				&-				&$<$71	\\
0849+27	&	$<$2.4 			&- 				&- 			\\
%1005+48 	&	5.6$\pm$1.0 		&- 				&-\\
1014+05	&	$<$3.0	 		&- 				&- 			\\
1102+11	&	$<$3.0 			&- 				&- 			\\
1159+01	&	-				&-				&$<$150 	\\
1159+06 	&	$<$6.0 			&- 				&$<$161 	\\
1229+09	&	$<$3.3 			&- 				&$<$66 	\\
1237+47	&	4.6$\pm$1.0 	&- 				&- 			\\
1304+13	&	$<$3.0 			&- 				&- 			\\
1327+03	&	$<$3.6 			&- 				&$<$201 	\\
1335+02	&	-				&-				&$<$78	\\
1337$-$02 &	-				&-				&$<$66	\\
1404+07	&	-				&-				&$<$90	\\
1406+34	&	9.3$\pm$0.8 	&- 				&- 			\\
\hline
\end{tabular}
\end{table}

\section{Discussion}
\label{sec:discussion}

In the following, we discuss our findings and put them in the context of the works present in the literature.

%%%%%%%
\subsection{Low-frequency components}

A consistent number of restarting radio sources have been found in the past 20 years, and hypotheses about their nature have been proposed (\citealt{Czerny, Wu, Marecki}). A fraction corresponds to GPS sources or CSS sources, known to be young radio sources. In some cases, up to three radio phases are detectable in the same object (\citealt{Brocksopp}). \cite{Czerny} associate the intermittent activity of the central engine with the radiation pressure instability of the accretion disk.
Concerning our results in the MHz range, the fact that hints of old, extended, radio components were found, despite the limited number of objects observed with the GMRT, could indicate that an age of $10^7$-$10^8$ years for the BAL-hosting QSOs is not rare. This finding is in line with what arose from our previous works (\citealt{Bruni,Bruni2}), where we discussed the presence of old components in the SED of our sample. In some cases a radio-restarting scenario, and the complex dynamics involved by that, could be invoked to explain a two-component SED (1159+01, 1159+06, 1335+02, 1406+34), showing both a GHz-peaked and a MHz component. In other cases (1014+04, 1229+09, 1304+13, 1327+03) a peak in the MHz range could still be present, but not seen because of insufficient frequency coverage. Considering this, and the SEDs of objects previously studied in \cite{Bruni}, a $\sim$70\% of our sample could be in a GPS or GPS+CSS phase.

Once again this shows how BAL-producing outflows can be present not only in young radio sources, but also in more complex scenarios. For example, sources found to have multiple radio phases ongoing (CSS+GPS), should have gone through an unstable radiation pressure phase, causing an intermittent BAL-producing outflow acceleration (favoured during the high-pressure phases). This could include young, just-started radio sources (GPS/HFP) as well as restarted radio sources (e.g. those with a CSS+GPS SED). If this link with ignition/recollimation of the radio jet would be confirmed, an outflow collimation to form a jet (as already proposed by \citealt{Elvis}) could be invoked to explain the BAL variability.

%%%%%%%
\subsection{Dust emission}
\label{sec:discussion:dust}
Several works in the past years, especially with the advent of the \emph{Herschel} space observatory, has constrained the dust emission properties in AGN, and verified the connection with star-formation rate. Here we provide a comparison between our results for BAL QSOs in the mm-band and what found by other authors.

A very similar study to ours, in terms of instrumentation, sensitivity, and setup, is the one presented by \cite{Omont}. They performed 250-GHz observations of 35 optically luminous RQ QSOs ($M_{B}<-$27.0), with a similar redshift range to ours (1.8$<z<$2.8), performed with the MAMBO bolometer at the same frequency. They found that 26$\pm$9\% of the sources present an emission at that frequency. %This fraction should be even higher for RL QSOs, since recent results of the {\emph{Herschel}}-ATLAS project indicate that powerful radio jets are associated with high star formation rates (\citealt{Kal}), and consequently should show stronger dust emission.
Since they reached an RMS very similar to our observations ($\sim$1 mJy) we can compare our detection rate with this percentage: only 1 out of 17 ($\sim$6\%) of our sources shows 250-GHz emission attributable to dust, a substantially smaller fraction than the one found by these authors.

More recently, \cite{Kal} published results of the {\emph{Herschel}}-ATLAS project regarding FIR properties of radio-loud and radio-quiet QSOs. Using five different bands (100, 160, 250, 350, 500 $\mu$m) they fitted the 
dust emission for both groups. At the mean redshift of our sample (z$\sim$2.3) they found a mean flux density for radio-loud QSOs of 23.5$\pm$2.1 mJy at 350 $\mu$m (corresponding to our 850 GHz observations), and 21.3$\pm$2.8 mJy at 500 $\mu$m ($\sim$600 GHz). These are values below the sensitivity we reached with APEX, but using a simple power law to extract the expected mean flux density at 250 GHz from the previous values, we obtain a value of $\sim$17 mJy, that would be well detectable with the RMS of $\sim$1 mJy we could reach at IRAM-30m. This could suggest that our sample of RL BAL QSOs is also poorer in dust content than normal RL QSOs, in addition to RQ ones (as seen comparing with \citealt{Omont}).

In the light of these works, our results support the idea of BAL QSOs not being specially dusty objects. In addition, although with modest statistical significance, our work suggests that dust emission in the RL BAL QSO sub-class could be weaker than expected.
\cite{Tombesi} found that fast outflows ($\sim$0.25c) from the accretion disk in AGN can hamper star formation, impacting the interstellar medium. BAL-producing outflows, although at a lower ionization stage than the ones considered by those authors, can present velocities up to 0.2c, and, in the light of these results, could have a similar effect on the SFR of the host galaxy. 
This needs to be investigated with further observations, that could confirm this on larger samples of RL BAL QSOs.

\section{Conclusions}

We performed observations in the m-band with the GMRT of 5 RL BAL QSOs, that already showed hints of emission in the MHz range, and in the mm-bands for 17 RL BAL QSOs, from our previously studied sample. We aimed at exploring the emission in the low-frequency regime, and the grey-body emission from dust, respectively. The conclusions are the following:

\begin{itemize}

\item All 5 objects observed at low frequencies present emission from extended components, indicating the presence of an old radio emission. In some cases a restarting radio-activity can be invoked to explain the double component, MHz and GHz-peaked, present in their SEDs. This could suggest an intermittent BAL phase, associated with periods of radio-restarting activity.
This is also supported by the morphology of some objects, both from this work (0849+27) and from \cite{Bruni2}, and by the fact that $\sim$70\% of our RL BAL QSOs sample can be considered in a GPS or CSS+GPS phase, thanks to the data presented here.

\item Only 1 out of 17 sources ($\sim$6\%) presents a clear contribution at 250 GHz from the dust grey-body  emission. In the other cases 3-$\sigma$ upper limits have been derived. Comparing our results with the fraction of dust-rich RQ QSOs found by \cite{Omont}, resulting in a percentage of $\sim$26\%, we found that RL BAL QSOs do not present a larger fraction of dust-rich objects with respect to the RQ QSO population. Also comparing with more recent works from the \emph{Herschel}-ATLAS collaboration (\citealt{Kal}), we found a lack of dust emission with respect to mean values for RL QSOs. Since the amount of dust can be connected to star-formation and thus to the age of the host galaxy, this results suggest that RL BAL QSOs are not a hosted by particularly young galaxies. The fact that, despite of radio loudness, these objects present  even less dust emission (and consequently a lower star formation rate) than RQ QSOs, could suggest that BAL-producing outflows are able to hamper star formation in the host galaxy.

\item Both the obtained results, from observations performed at m- and mm-wavelengths, suggest that BAL QSOs
 are not commonly young radio objects, or objects still expelling their dust cocoon from the central region. They could be radio-restarting objects, which present relativistic outflows in conjunction with some periods of favourable emission/acceleration conditions. 

\end{itemize}

%%%%%%%%%%%%%%%%%%%%%%%%%%%%%%%%%%%%%%%%%%%%%%%%%%%%%%%%%%%%%%%%%%%%%%%%%%%%%%%%%%

\begin{acknowledgements}
Part of this work was supported by a grant of the Italian Programme for
Research of National Interest (PRIN No. 18/2007, PI: K.-H. Mack)
The authors acknowledge financial support from the Spanish Ministerio de Ciencia e Innovaci\'on under project 
AYA2008-06311-C02-02.
We thank the staff of the GMRT that made these observations possible. GMRT is run by the National Centre for Radio Astrophysics of the Tata Institute of Fundamental Research.
This publication is based on data acquired with the Atacama Pathfinder Experiment (APEX). APEX is a collaboration between the Max-Planck-Institut f\"ur Radioastronomie, the European Southern Observatory, and the Onsala Space Observatory.
This work is partly based on observations carried out with the IRAM-30m Telescope. IRAM is supported by INSU/CNRS (France), MPG (Germany) and IGN (Spain).
\end{acknowledgements}

%-------------------------------------------------------------------

\end{document}